\documentclass[12pt]{article}
\usepackage{cmb,epsfig}
\psfull

\newcommand{\ea}{{\it et~al.\/} }
\newcommand{\dg}{\nobreak^\circ}
\newcommand{\df}{dec $+40\dg$ }

\newcommand{\ho}{\mbox{$\mbox{H}_0$} }
\newcommand{\simlt}{\mbox{$\stackrel{<}{_{\sim}}$} }
\newcommand{\simgt}{\mbox{$\stackrel{>}{_{\sim}}$} }

\newcommand{\ie}{{\it i.e.\ }}
\newcommand{\eg}{{\it e.g.\ }}

\newcommand{\qrms}{\mbox{$Q_{rms-ps}$} }

\newcommand{\be}{\begin{equation}}
\newcommand{\ee}{\end{equation}}

\begin{document}
\heading{A FIRST DETERMINATION OF THE POSITION
OF THE `DOPPLER' PEAK}

\author{Stephen Hancock and Graca Rocha}{MRAO, Cambridge,UK.}{$\,$}

\begin{abstract}
Cosmological theories for the origin and evolution of structure in the
Universe are highly predictive of the form of the angular power spectrum of
cosmic microwave background fluctuations.  We present new results from a
comprehensive study of CMB observations which provide the first measurements
of the power spectrum all the way down to angular scales of $\sim 10$
arcminutes. On large scales a joint likelihood analysis of the COBE and
Tenerife data fixes the power spectrum normalisation to be $\qrms = 21.0\pm
1.6 \mu$K for an initially scale invariant spectrum of fluctuations. The
combined data are consistent with this hypothesis, placing a limit of $n=1.3
\pm 0.3$ on the spectral slope.  On intermediate scales we find clear evidence
for a `Doppler' peak in the power spectrum localised in both angular scale and
amplitude.  This first estimate of the angular position of the peak is used to
place a new direct limit on the curvature of the Universe, corresponding to a
density of $\Omega=0.7^{+1.0}_{-0.4}$.  Very low density open Universe models
are inconsistent with this limit unless there is a significant contribution
from a cosmological constant.
\end{abstract}

\section{Introduction}
Observations of the Cosmic Microwave Background (CMB) radiation provide
information about epochs and physical scales that are inaccessible to
conventional astronomy. In contrast to traditional methods of determining
cosmological parameters, which rely on the combination of results from local
observations\cite{os}, CMB observations provide direct
measurements\cite{be87,white} over cosmological scales, thereby avoiding the
systematic uncertainties and biases associated with conventional techniques.
Assuming that the fluctuations conform to a random Gaussian field, then all of
the statistical properties are contained in the angular power spectrum and
consequently tracing out the form of the CMB power spectrum is a key goal of
observational cosmology\cite{be87,bond1,bond2,scott,hu, kami,rhatra}.  CMB
observations on different angular scales are probing different physical
effects (see White \ea 1995 \cite{white} and Scott \ea 1995 \cite{scott} for
comprehensive reviews).  On scales $\simgt 2\dg$, which is the size of the
horizon at last scattering of the CMB photons, the gravitational redshifting
of the CMB photons through the Sachs-Wolfe effect is the dominant process.
The presence of either a background of primordial gravity waves and/or mass
fluctuations at recombination, would lead to fluctuations in the CMB. The
scalar mass density fluctuations lie outside the horizon and are predicted by
inflationary theory to have a scale-invariant flat power spectrum \ie $n=1$.
In Section 2 we use the combined large scale COBE and Tenerife observations to
delimit the spectral slope $n$ and to fix the power spectrum normalisation
$\qrms$.

On scales of $\sim 0.2\dg-2.0\dg$ the scattering of the CMB photons
during acoustic oscillations of
the photon-baryon fluid at recombination \cite{cdm2}
is expected to imprint
characteristic `large' amplitude 
peaks into the CMB power spectrum. The position $l_p$ of the main peak 
reflects the size of the horizon at last scattering of the CMB photons and is
determined almost entirely by the geometry of the Universe.
As a result one finds that \cite{hu,kamio} $l_p$
depends directly 
on the density of the Universe according to $l_p \propto 1/\sqrt{\Omega}$.
The height of the peak is directly proportional
to the fractional mass in baryons $\Omega_b$ and also varies according
to the expansion rate of the Universe as specified by the Hubble constant
$\mbox{H}_0$; in general \cite{hu}
for baryon fractions $\Omega_b \simlt 0.05$, increasing \ho
reduces the peak height whilst the converse is true at higher baryon densities.
Consequently by measuring the amplitude of the intermediate
scale CMB  fluctuations relative to the CMB fluctuations resulting from
scalar density
perturbations on large scales
we can trace out the CMB power spectrum
and hence directly estimate $\Omega$,
$\Omega_b$ and \ho from the position and amplitude of the main peak.
In Section 3, we use current CMB observations, including new
data from the CAT, Tenerife and COBE experiments to build up a conservative
and consistent picture of the CMB power spectrum on large and intermediate
scales and hence to obtain a first estimate of both the position and
amplitude of the Doppler
peak.

\section{Joint likelihood analysis of COBE and Tenerife observations}
Both the COBE satellite observations \cite{bennett} and the ground-based
observations from the Tenerife experiments \cite{davies}
are on sufficiently large angular scales that they probe fluctuations
that are beyond the horizon scale at recombination and are hence
still in the linear growth regime. Such data can
therefore be used to investigate the spectral slope of the initial
primordial fluctuation spectrum generated in the early Universe.
Numerous attempts \cite{smoot,me96,nature94,tb95,bennett,gorski}
have been made to determine both the slope and
normalisation of the power spectrum at small multipoles. The approach 
detailed in Hancock \ea \cite{me96} and reviewed here differs from
previous analyses in that for the first time the COBE
and Tenerife data have been used {\em together} taking into account
{\em all} correlations between the two data sets. This method, developed
in collaboration with Max Tegmark, uses a direct brute force
calculation 
of the likelihood function for the combined data. 

We apply the likelihood analysis to the COBE two-year \cite{cobetwo} data and the
Tenerife \df scan \cite{me96,nature94},
assuming a power law model with free parameters
$n$ and $\qrms$. The COBE Galaxy-cut two-year map consists of 4038
pixels, whilst the Tenerife Galaxy cut (RA $161\dg-230\dg$) scan contains 70
pixels, requiring a 4108 $\times$ 4108 covariance matrix for a joint
likelihood analysis of the data.
We arrange the pixels in a 4108-dimensional vector ${\bf \Delta  T}=
(\Delta T_1, \Delta T_2,...\Delta T_{4108})$
and compute the 
likelihood function $L(n,\qrms) \propto \exp -({\bf \Delta T^T V^{-1} \Delta T})$
as in Tegmark and Bunn 1995 \cite{tb95} by Cholesky
decomposition of the 4108 $\times$ 4108 covariance matrix ${\bf V}$
at a dense grid
of points in the $(n,\qrms)$-parameter space, marginalizing over
the four ``nuisance parameters" that describe the monopole and dipole.
The covariance matrix consists of three parts: a $70 \times 70$ block with the
covariance between the Tenerife pixels, a $4038 \times 4038$ 
block with the covariance between the COBE pixels,
and off-diagonal $4038 \times 70$ blocks containing the covariance between 
the Tenerife and COBE pixels. 
In this way, we fully
account for intrinsic correlations due to the CMB structure and
correlations due to sampling with the different instruments.
Additionally in forming the likelihood function we have intrinsically
incorporated the effects of cosmic and sample variance for the two
data sets, plus
random noise and the interdependence of the model parameters.

The resulting normalised joint likelihood function depicted in Figure 4
of Hancock \ea (submitted \cite{me96})
thus provides an accurate description of the constraints placed
on $n$ and \qrms by the joint data set. The likelihood is seen to
peak at $n=1.37$,
$\qrms=16.1\mu$K, with a 68\% confidence region (uniform prior)
encompassing 0.90 to 1.73 in $n$ for
$\qrms$ in the range from $12.1$ to $22.9$ $\mu$K. Margenalising over \qrms
one finds $n=1.3\pm 0.3$, whilst conditioning on
$n=1$, one finds a power spectrum normalisation of $\qrms=21.0\pm 1.6$.
These results using COBE 2-year and Tenerife \df data
are comparable to those obtained using the COBE 4-year
data \cite{bennett}, for which $n=1.2 \pm 0.3$ and $\qrms=18\pm1.6 \mu$K for $n=1$. The joint analysis of the COBE 4-year data plus a significantly
extended Tenerife sky area is in progress and is expected to
improve on these limits. However, it is clear that the current Tenerife
and COBE results offer a consistent picture on large scales and do
not favour values of $n$ less than unity.
In the case of power law inflation, such large
values of $n$ do not allow for a significant contribution from tensor
modes, giving us confidence in normalising the scalar power spectrum
to the large scale anisotropy data. In the following, we will {\em assume}
that this is the case and will proceed to compare the large and
intermediate scale anisotropy measurements to test for the presence
of a Doppler peak.

\section{The CMB Power Spectrum}
Reconstructing the CMB power spectrum over large and intermediate
angular scales requires the simultaneous use of data from a number
of different experiments, all with their own classes of uncertainties.
At the time of writing, there are numerous CMB experiments operating
worldwide, and it is appropriate here to restrict ourselves to the
subset of experiments which have produced conclusive evidence for
the detection of CMB anisotropy.
Clear detections have
now been reported by a number of different groups,
using observations from satellites (COBE\cite{smoot,bennett}),
ground-based switching experiments (Tenerife\cite{me96},\cite{nature94},
Python\cite{python}, South Pole\cite{spole}, 
Saskatoon\cite{sask}),
balloon mounted instruments (ARGO\cite{argo}, MAX\cite{max}, MSAM\cite{msam1,msam2})
and more recently ground-based interferometer telescopes (CAT\cite{cat}).
Given the difficulties inherent in observing CMB anisotropy, it is possible
and indeed likely,
that some of these results are contaminated by foreground effects\cite{gr}.
Determining the form of the CMB power spectrum in order to trace out the
Doppler peak requires a careful, in-depth
consideration of the CMB measurements from the different experiments
within a common framework as presented in Hancock \ea (submitted \cite{menew}); the full
details including a discussion of foreground contamination
are presented in Rocha \ea (in preparation\cite{gr}). In this paper and the
following contribution \cite{gs}, we present
our principal findings.
We consider all of the latest CMB measurements, including new results
from COBE, Tenerife, MAX, Saskatoon and CAT, with the exception of
the MSAM results and the
MAX detection in the Mu Pegasi region which
is contaminated by dust emission\cite{fischer}.
 On the largest scales corresponding to small $l$, new COBE
\cite{bennett} and Tenerife
\cite{me96} results improve the power spectrum normalisation, whilst
significant gains in knowledge at high $l$ are provided by new
results from the Saskatoon and CAT
experiments. The full data set spans a range of $2$ to $\sim 700$ 
in $l$, sufficient to test for the main Doppler peak out to $\Omega =0.1$.

\subsection{The flat bandpower approximation}
\begin{figure}
\centerline{\epsfig{figure=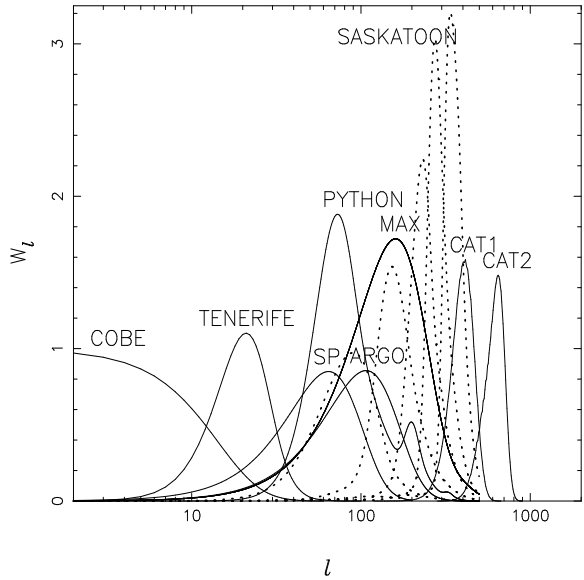,height=3.0in,width=3.0in,angle=-90}}
\caption{The experimental window functions $W_l$.}
\end{figure}
The competing models for the origin and evolution of structure
predict\cite{be87,hu},
the shape and amplitude of the CMB power
spectrum and its Fourier equivalent, the autocorrelation function
$C(\theta) =<\Delta T({\bf n_1})\Delta T({\bf n_2})>$ where ${\bf n_1}
\cdot {\bf n_2} =\cos \theta$.
Expanding the intrinsic angular correlation function $C(\theta)$ in terms
of spherical harmonics one obtains:
$C(\theta)= \sum_{l\ge2}^{\infty} (2l+1) C_l P_l(\cos (\theta)) /4 \pi$,
where low order multipoles $l$ correspond to large angular scales $\theta$
and large $l$-modes are equivalent to small angles on the sky.
The different experiments sample different $l$-modes
according to their window functions $W_l$,
as shown in Figure 1 :
for a detailed discussion
of window functions see \cite{whitewindow1,whitewindow2}.
The observed power in CMB fluctuations as seen through a window $W_l$
is given by
\be
C_{obs}(0)=\left (\frac{\Delta T_{obs}}{T} \right )^2 = \sum_{l \ge 2}^{
\infty} (2l+1) C_l W_l/4 \pi.
\label{eq:cobs}
\ee 
The $C_l$ are predicted by the cosmological theories
and contain all of the relevant statistical
information for models described by Gaussian random fields\cite{be87}.   
Given $W_l$, then for the $C_l$'s corresponding to the theoretical model under 
consideration   
it is possible to obtain the value of $\Delta T_{obs}$
one would expect to observe using the chosen experiment.
This value can then be compared to
the value actually observed to test the cosmological model.

We take the reported CMB detections and convert them to a common framework
of flat bandpower results\cite{bond1,bond2} as given in Table 1 of the contribution by Rocha and Hancock (this volume).
In order to use the observed anisotropy levels to place constraints on the CMB
power spectrum one must in general know the form of the $C_l$ under test.
However, in most cases the form of $C_l$ can be represented by a flat
spectrum $C_l \propto C_2/(l(l+1))$ over the width of a given experimental window,
so that the bandpower is $\Delta T_l/T=\sqrt{C_{obs}(0)/I(W_l)}$, where
we define $I(W_l)$ according to Bond (1995)\cite{bond1,bond2} as
$I(W_l)=\sum_{l=2}^{\infty} (l+0.5)W_l/(l(l+1)$. This bandpower estimate is
centred on the effective multipole $l_e=I(lW_l)/I(W_l)$.
In many instances experimenters now report results directly for a flat
spectrum and when this is not so
we have converted the quoted power in fluctuations into
the equivalent flat band estimate.
Each group has obtained limits on the intrinsic anisotropy
level using a likelihood analysis (see \eg Hancock \ea 1994\cite{nature94}), which
incorporates uncertainties due to random errors, sampling variance\cite{samvariance} and
cosmic variance\cite{cosvariance1,cosvariance2}.
The form of the likelihood function is not necessarily Gaussian,
and strictly one requires a method that will utilise the full
likelihood functions from all experiments consistently. However,
given the relatively large error bars on most of the reported data points
it is sufficient for our purposes here to approximate all likelihood
results as originating from a Gaussian distribution, giving the
one-sigma error bounds in column three of Table 1 (see \cite{gs}) by 
averaging the
difference in the reported 68\% upper and lower limits and the best fit
$\Delta T_l$. This bias introduced by this averaging is discussed in \cite{gr}.

Results from the MSAM experiment are not included
here, because they do not provide an independent measure of the power spectrum
since their angular sensitivity and sky coverage are already incorporated
within the Saskatoon measurements. Netterfield \ea\ \cite{sask} report good
agreement between the MSAM double difference results and Saskatoon
measurements, although the discrepancy with the MSAM single difference data is
yet to be resolved.

\subsection{Estimating the parameters of the Doppler Peak}
\begin{figure}
\centerline{\epsfig{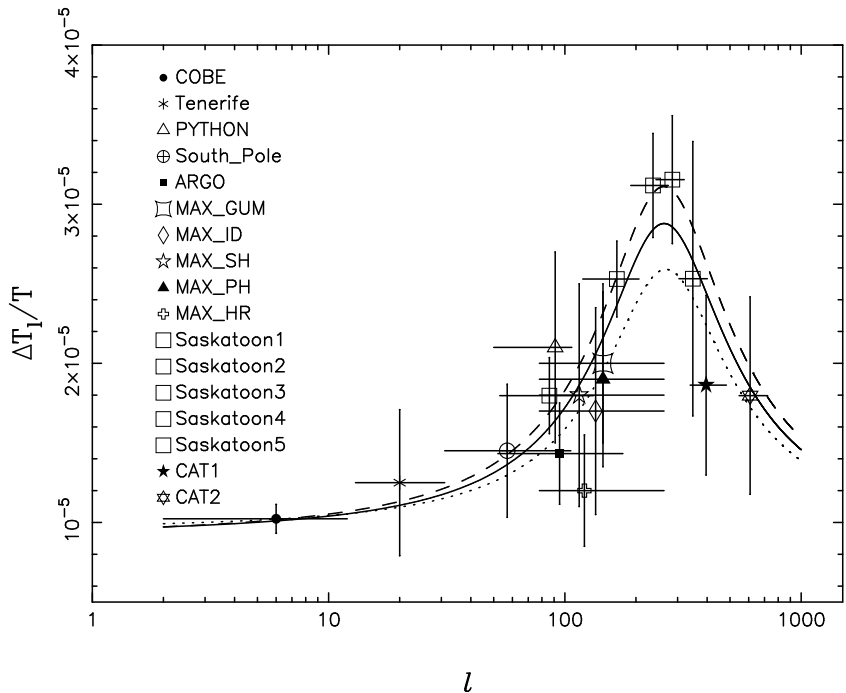}}
\caption{The data points from Table 1 (see \cite{gs}) are shown compared to the best fit
analytical CDM model. The dotted and dashed lines show the best fit models
which are obtained when the Saskatoon calibration is adjusted by $\pm 14\%$. The
data points from the MAX experiment are shown offset in $l$ for clarity \label{fig:fig2a}}
\end{figure}

The data points from Table 1 (see \cite{gs}) are plotted in Figure~\ref{fig:fig2a},
in which the horizontal bars represent the range of $l$ contributing to each
data point. There is a noticeable rise in the observed power spectrum at
$l\simeq 200$, followed by a fall at higher $l$, tracing out a clearly defined
peak in the spectrum.  In the past several groups \cite{scott,kami,rhatra} have
attempted to determine the presence of a Doppler peak, but only now are the
data sufficient to make a first detection and to put constraints on the closure
parameter $\Omega$.  As a first step, we adopt a simple three parameter model
of the power spectrum, which we find adequately accounts
for the properties of the principal Doppler peak for both standard Cold Dark
Matter (CDM)  models \cite{cdm1,cdm2} and open Universe ($\Omega<1$) models
\cite{kami}. The functional form chosen is a modified version of that used in 
Scott, Silk \& White \cite{scott} --- we choose the following: 
\be
l(l+1)C_l=6C_2\left(1+\frac{A_{peak}}{1+y(l)^2}\right ) {\huge /}
\left(1+\frac{A_{peak}}{1+y(2)^2}\right )
\label{eq:peak}
\ee 
where $y(l)=(\log_{10}l-log_{10}(220/\sqrt{\Omega}))/0.266$. In this
representation $C_2$ specifies the power spectrum normalisation, whilst the
first Doppler peak has height $A_{peak}$ above $C_2$, width $\log_{10}l=0.266$
and for $\Omega=1.0$ is centred at $l\simeq 220$.  By appropriately specifying
the parameters $C_2$, $A_{peak}$ and $\Omega$ it is possible to reproduce to a
good approximation the $C_l$ spectra corresponding to standard models of
structure formation with different values of $\Omega$, $\Omega_b$ and
$\mbox{H}_0$. Such a form will not reproduce the structure of the {\em
secondary\/} Doppler peaks, but we have checked the model against the overall
form of the $\Omega =1$ models of Efstathiou and the open models reported in
Kamionkowsky \ea\ \cite{kami} and find that this form adequately reflects
the properties of the main peak. This satisfies our present considerations
since the current CMB data are not yet up to the task of discriminating the
secondary peaks.  Varying the three model parameters in
equation~(\ref{eq:peak}) we form $C_l$ spectra corresponding to a range of
cosmological models, which are then used in equation~(\ref{eq:cobs}) to obtain
a simulated observation for the $i$th experiment, before converting to the
bandpower equivalent result $\Delta T_l[C_2,A_{peak},\Omega](i)$.  The
chi-squared for this set of parameters is given by
\begin{displaymath}
\chi^2(C_2,A_{peak},\Omega)=\sum_{i=1}^{nd}
\frac{(\Delta T_l^{obs}(i) -\Delta T_l[C_2,A_{peak},
\Omega](i))^2}{\sigma_i^2}, 
\label{eq:chi}
\end{displaymath}
for the $nd$ data points in Table~1 (see \cite{gs}) and the relative likelihood function is
formed according to $L(C_2,A_{peak},\Omega) \propto \exp( -
\chi^2(C_2,A_{peak},\Omega)/2)$.  We vary the power spectrum normalisation
$C_2$ within the 95 \% limits for the COBE 4-year data \cite{bennett} and
consider $A_{peak}$ in the range 0 to 30 and values of the density parameter up
to $\Omega=5$.  The data included in the fit are those from Table 1 (see \cite{gs}), which with
the exception of Saskatoon include uncertainties in the overall calibration.
There is a $\pm 14$\% calibration error in the Saskatoon data, but since the
Saskatoon points are not independent this will apply equally to all five points
\cite{sask}.  The likelihood function is evaluated for three cases: (i) that the
calibration is correct, (ii) the calibration is the lowest allowed value and
(iii) the calibration is the maximum allowed value.  In each case the likelihood
function is marginalised over $C_2$ before calculating limits on the remaining
two parameters according to Bayesian integration with a uniform prior.

\section{Results and Discussion}

\begin{figure}
\centerline{\epsfig{figure=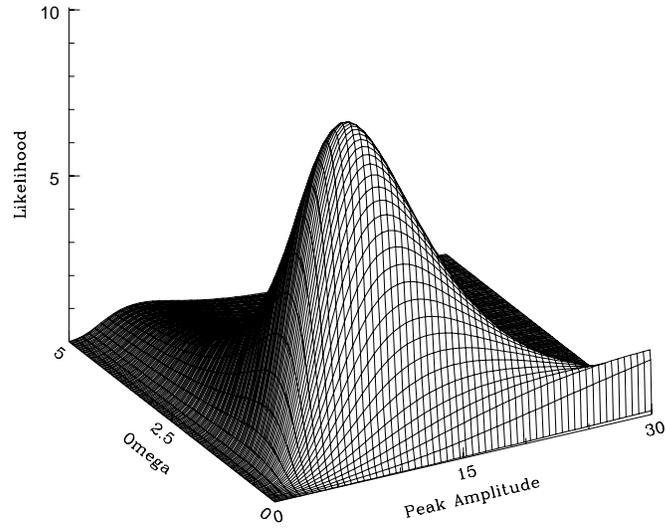,width=4.0in,angle=0}}
\caption{The likelihood surface for $\Omega$ and $A_{peak}$. (The nominal
Saskatoon calibration is assumed.) \label{fig:3d}}
\end{figure}

\begin{figure}
\centerline{\epsfig{figure=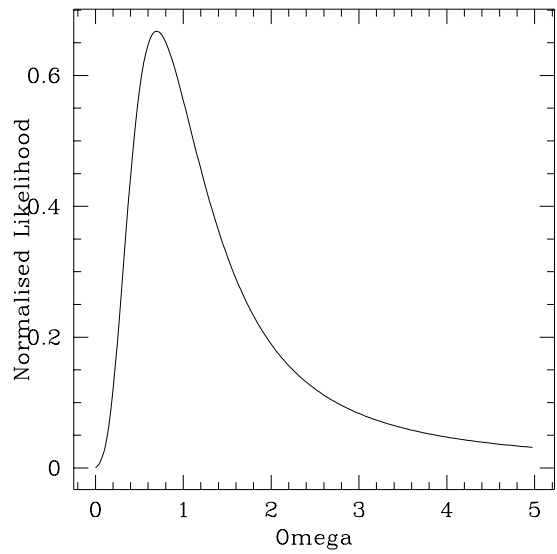,height=3.0in,width=3.0in,angle=0}}
\caption{The 1-D conditional likelihood curve for $\Omega$. \label{fig:omega2}}
\end{figure}

%We now perform a likelihood fit to the model power
%spectrum.
In Fig.~\ref{fig:3d}
the likelihood function obtained from fitting the model $C_l$ spectra to the
data of Table~1 (see \cite{menew}) is shown plotted as a function of the amplitude and position
($\Omega$) of the Doppler peak. The highly peaked nature of the likelihood
function in Fig.~\ref{fig:3d} is good evidence for the presence of a Doppler peak
localised in both position ($\Omega$) and amplitude. In Fig.~\ref{fig:omega2}
we show the 1-D conditional likelihood curve for $\Omega$, obtained by cutting
through the surface shown in Fig.~\ref{fig:3d} at the best-fit value of $A_{peak}$.
The best fit value of $\Omega$ is 0.7 with an allowed 68\%
range of $0.30 \le \Omega \le 1.73$.

In Figure~\ref{fig:fig2a} the
best fit model, represented by the solid line, is shown compared to the data
points, assuming no error in the calibration of the Saskatoon observations.
The chi-squared per degree of freedom for this model is $0.9$, implying a good
fit to the data.  The peak lies at $l=263_{-94}^{+139}$ corresponding to a
density parameter $\Omega =0.70_{-0.4}^{+1.0}$; the height of the peak is
$A_{peak}=9.0_{-2.5}^{+4.5}$.  The dashed and dotted lines show the best fit
models ($\Omega =0.70_{-0.37}^{+0.92}$, $A_{peak}=11.0_{-4.0}^{+5.0}$ and
$\Omega =0.68_{-0.4}^{+1.2}$, $A_{peak}=6.5_{-2.0}^{+3.5}$ respectively)
assuming that the Saskatoon observations lie at the upper and lower end of the
permitted range in calibration error.  
These likelihood results using the analytic form for the $C_l$ and the results from a more detailed
analysis using exact models (see Hancock \ea submitted \cite{menew}; Rocha and Hancock, this volume) imply that independent of
calibration uncertainties in the data, current CMB data are inconsistent with
cosmological models with $\Omega < 0.3$.

\section{Conclusions}

Our current results provide good
evidence for the Doppler peak, verifying a crucial prediction of cosmological
models and providing an interesting new measurement of fundamental cosmological
parameters. In Rocha \ea\ \cite{gr},
a detailed comparison of the CMB data is made with the theoretical power
spectra predicted by a range of flat, tilted, reionized, open models and models
with non-zero cosmological constant.
The existence of the Doppler peak has important consequences for
the future of CMB astronomy, implying that our basic theory is correct and that
improving our constraints on cosmological parameters is simply a matter of
improved instrumental sensitivity and ability to separate out foregrounds.  New
instruments such as VSA \cite{lasenbyandhancock}, MAP and the proposed
COBRAS/SAMBA satellite \cite{cobras} will provide this improved sensitivity and
should delimit $\Omega$ and other parameters with unprecedented precision.

\acknowledgements{Thanks to all the members of 
the CAT and Tenerife teams for their help and assistance in this work.
and to B. Netterfield for supplying
the Saskatoon window functions.
S. Hancock wishes to acknowledge a Research
Fellowship at St.\ John's College, Cambridge, U.K.
}

\vfill
\end{document}